# Sinogram interpolation for sparse-view micro-CT with deep learning neural network

Xu Dong[a], Swapnil Vekhande[b], Guohua Cao[a*]
[a]Virginia Tech-Wake Forest University School of Biomedical Engineering and Sciences, Virginia Polytechnic Institute and State University, Blacksburg, VA, USA, 24061; [b]Department of Electrical and Computer Engineering, Virginia Polytechnic Institute and State University, Blacksburg, VA, USA, 24061

## ABSTRACT

In sparse-view Computed Tomography (CT), only a small number of projection images are taken around the object, and sinogram interpolation method has a significant impact on final image quality. When the amount of sparsity - the amount of missing views in sinogram data – is not high, conventional interpolation methods have yielded good results. When the amount of sparsity is high, more advanced sinogram interpolation methods are needed. Recently, several deep learning (DL) based sinogram interpolation methods have been proposed. However, those DL-based methods have mostly tested so far on computer simulated sinogram data rather experimentally acquired sinogram data. In this study, we developed a sinogram interpolation method for sparse-view micro-CT based on the combination of U-Net and residual learning. We applied the method to sinogram data obtained from sparse-view micro-CT experiments, where the sparsity reached 90%. The interpolated sinogram by the DL neural network was fed to FBP algorithm for reconstruction. The result shows that both RMSE and SSIM of CT image are greatly improved. The experimental results demonstrate that this sinogram interpolation method produce significantly better results over standard linear interpolation methods when the sinogram data are extremely sparse.

## 1. INTRODUCTION

CT allows physicians to diagnose injuries and disease more quickly and accurately than other imaging modalities. The useful features that distinguish CT are fast scanning and spatial resolution. However, there has been increase in incidence of radiation-induced carcinogenesis. CT alone contributes almost one half of the total radiation exposure from medical use and one quarter of the average radiation exposure per capita in the USA [1].

Radiation dose is one of the most significant factors determining CT image quality and thereby the diagnostic accuracy and the outcome of a CT examination. It should only be reduced under the condition that the diagnostic image quality is not sacrificed. The method like CT system optimization, ccan range minimization and automatic exposure control have yielded some benefit. CT imaging with sparsely sample projections is another effective approach to reduce radiation dose. The main challenge for sparse-view CT is its image reconstruction.

Image reconstruction from sparsely sampled signal falls under ill–posed inverse problem and has been addressed in broadly two ways. Compressive sensing theory has been applied to the reconstruction problem in the form of simultaneous algebraic reconstruction technique and its variations [2, 3]. Total variation loss minimization [4, 5] using an optimization solver framework was another effective solution. Such iterative approaches are complex in terms of computations. On the other hand, filtered back projection (FBP) remains one of the gold standards for reconstruction.

Research was carried out to synthesize missing sinogram data before feeding it to the FBP to reduce artifacts in reconstructed images. Typically, such sinogram synthesis task was handled with linear interpolation or nearest-neighbor interpolation. However, those conventional techniques are not adequate and still exhibit significant image artifact after reconstruction. This is especially true when a large percentage of sinogram data need to be interpolated. Recently, deep learning (DL) has been successfully used in numerous image processing applications [6]. Similarly, convolutional neural networks [7-9] have also been applied to address the sparse-view CT problem. Another type is the generative adversarial networks-based methods [10, 11]. There have been efforts to develop end-to-end deep learning framework [12-14] but generalization of such models remains to be tested for all applicable cases. There are also other methods to address the angular resolution issue in sparse-view CT [15, 16]. Recently, Cho *et. al.* [15, 17] has proposed local linear interpolation on simulated data, and then residual UNET was used to synthesize missing data. However, so far most of those sinogram inpainting studies for sparse-view CT have been only tested with simulated sinogram data, and comparisons for the performance of those methods were mostly carried in the image domain rather than the sinogram domain, presumably due to the lack of the experimental sinogram as ground truth.

In this paper, our objective is to achieve sinogram inpainting on super-sparsely sampled experimental sinogram data. The sparsity (defined as the percentage of missing sinogram data) is as high as 90%. The realistic experimental sinogram data, due to their inherent imperfections from the complex imaging physics, detector response, as well as scanning trajectory, makes the problem a challenging task.

The paper is organized as follows. After the introduction, in section II we explained our methods for sinogram inpainting. We then discuss results in section III, and the discussions and conclusions are in section IV.

# 2. METHODS

## 2.1 Overview of the method

Our proposed method has two steps. The sparse-view sinogram will first be up-sampled by a simple linear interpolation algorithm. Then the sinogram is processed by a pre-trained deep learning neural network to remove the artifacts and enhance the image quality. The workflow of the proposed method is illustrated in Fig. 1. The network has to be trained on a training dataset in a supervised fashion. The DL synthesized sinogram is then fed to FBP algorithm for image reconstruction.

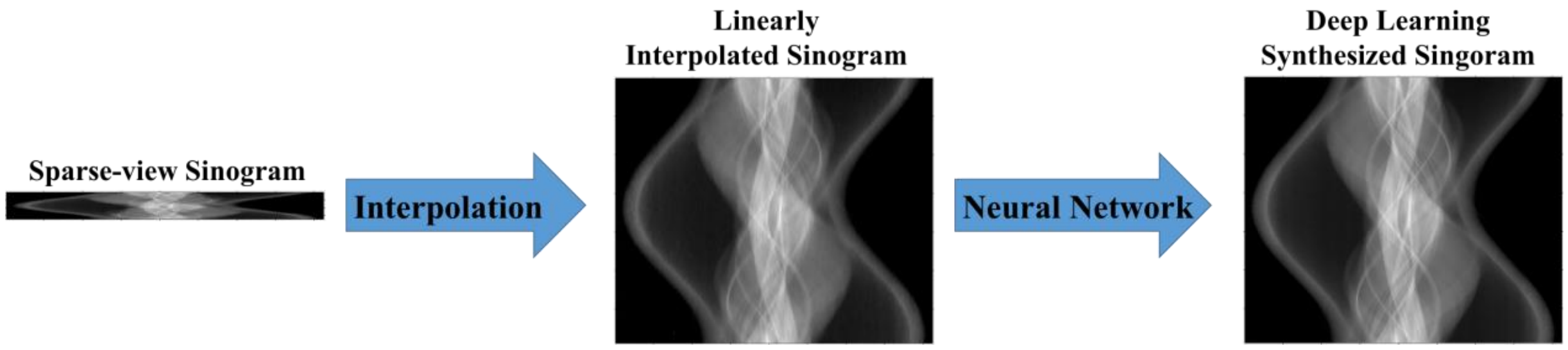

Figure 1. The workflow of the proposed sinogram synthesis method. The DL synthesized sinogram is then used for FBP reconstruction.

## 2.2 The architecture of proposed neural network

The neural network used for sinogram enhancement was designed based on the combination of U-Net and residual learning. The U-Net structure is one the most effective network structures for medical image processing [18]. We modified the original U-Net design in two folds. First, the max pool layers were replaced by convolutional layers with the stride of 2 by 2. Such replacement follows the suggestion made by Radford *et. al.* [19]. Second, the deconvolution is used to upscale the image in our design because of its good ability to restore the image features. Residual learning technique was utilized by adding the input image to the output of U-Net and taking the summation to be the eventual network output. The residual learning forces the network to find the difference between the input data and the ground truth data rather than to map the input image to ground truth image directly. This technique can help the network to learn faster and more effectively.

Instead of processing one sinogram as the whole, the proposed network process one image patch in size of 64 by 64 each time. Processing small image patches can highly reduce the computational cost and make the approach more feasible to be used in common computers. The network architecture and the structural parameters are illustrated in Fig. 2.

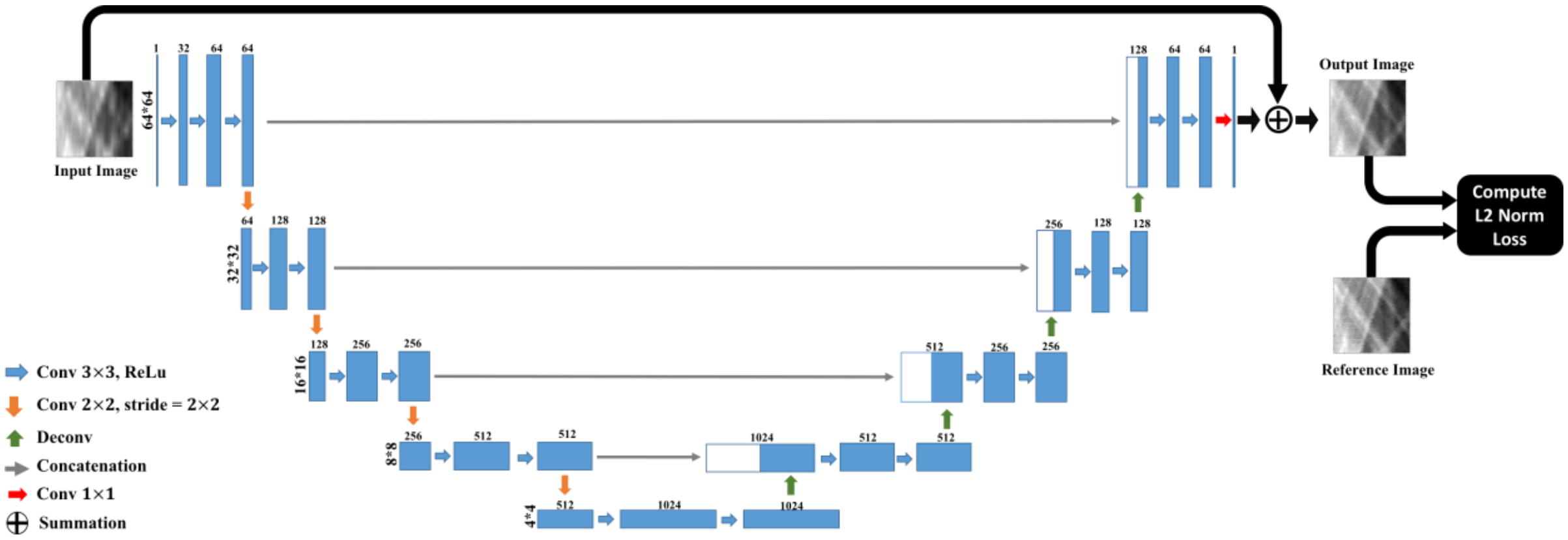

Figure 2. The architecture of the proposed neural network.

### 2.3 Data preparation

The sinograms were collected from a rat bone scan. The scan was conducted in a micro-CT (Xradia MicroXCT-200). The scanning protocol is reported in Table 1. While one scan can yield 1024 sinograms in total, only 260 sinograms in the central slices were only used in the study. We also cropped the projection data by cutting out 100 peripheral detector pixels at each side. Those pre-processing steps were applied to make sure all the collected sinograms are in good quality.

The sparse-view sinograms were constructed by selecting one projection in every 10 views. In contrast, the full-view sinograms were constructed by using all the projections. As a result, each sparse-view sinogram is in the size of 69 by 824 and each full-view sinogram is in the size of 688 by 824. We interpolated the sparse-view sinogram to upscale its size to be 688 by 824 by a linear interpolation algorithm. Then both interpolated sinograms and full-view sinograms were divided into 64 by 64 patches so that they can be feed into the neural network.

We randomly divided the 260 sinograms into training dataset (156 sinograms), validation dataset (13 sinograms), and test dataset (91 sinograms), which accounts for 60%, 5%, and 35% of the whole dataset respectively.

Table 1. The scanning protocol of micro-CT.

| | |
|---|---|
| **X-ray Source** | 60 kV, 8 W |
| **Source to isocenter distance** | 70 $mm$ |
| **Source to detector distance** | 134.56 $mm$ |
| **Detector pixel size** | $0.01 \times 0.01\ mm^2$ |
| **Detector resolution** | $1024 \times 1024$ |
| **Angular sampling** | 1 projection per 0.5 degree |
| **Angular coverage** | 344 degrees |
| **Exposure time** | 8 sec. per projection |

### 2.4 Network training

The interpolated sinograms were feed into the network, then the output would be compared with the corresponding full-view sinograms. The Euclidean distance between the neural network output and the full-view sinogram was computed as the loss function. After feeding all the training data into the network, the network parameters were updated with the objective to minimize the loss function. The process was repeated hundreds of times until the loss cannot be further optimized.

In our study, the Adam optimizer was used to optimize the loss. The learning rate was set to be 0.0002 initially and to decay 0.95 every 2000 steps. The algorithm was implemented in Python based on TensorFlow. The training was done on a workstation with Intel Core i5-7400 CPU and 16GB RAM. A GPU card (Nvidia GTX Titan X) accelerated the training process. All the convolution and deconvolution filters were initialized with random Gaussian distributions with zero mean and 0.01 standard deviation.

### 2.5 Evaluation

After the neural network being trained on training dataset, it was then applied to synthesize the sparse-view sinograms in the test dataset. The deep learning synthesized sinograms and the interpolated sinograms were compared with full-view sinograms (as shown in Fig. 3).

The sparse-view sinograms, interpolated sinograms, deep learning synthesized sinograms, and the full-view sinogram were reconstructed separately by Filtered Backpropagation algorithm. The reconstructed images are displayed in Fig 4. The images reconstructed from full-view sinograms were used as the reference for comparison. The root mean square error (RMSE) and structural similarity (SSIM) were calculated when comparing the images reconstructed from other approaches with reference images. The result is reported in Table 3.

## 3. RESULT

### 3.1 The evaluation of sinogram

The sparse-view sinogram, linearly interpolated sinogram, deep learning synthesized sinogram, and full-view sinogram (reference) are shown in Fig. 3. When compared with reference, the RMSE were calculated for linearly interpolated sinogram and deep learning synthesized sinogram. The result is displayed in Table 2. We can see the RMSE is improved from 0.008087 to 0.004709 after the sinogram is synthesized by the deep learning approach.

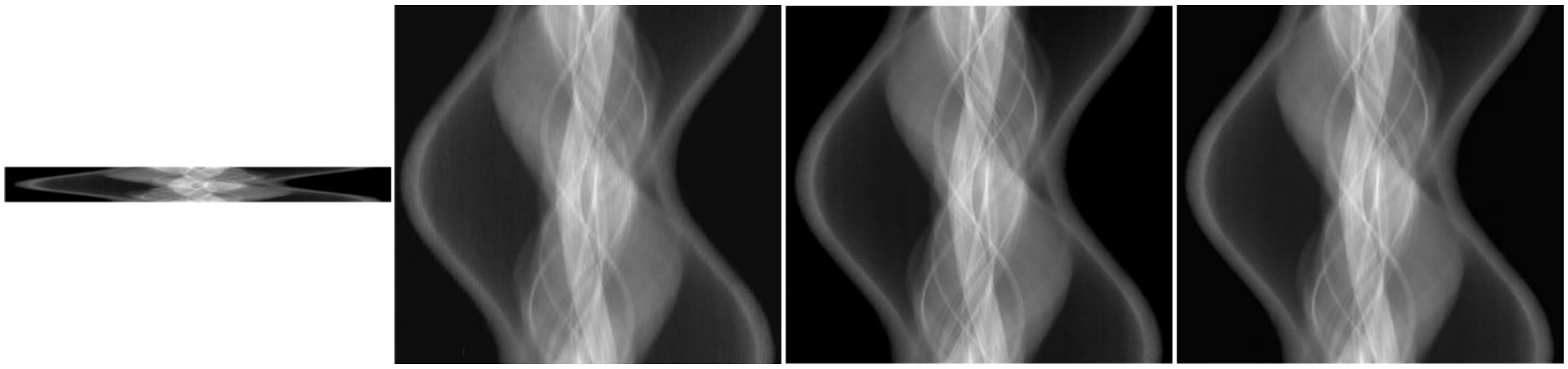

Figure 3. The sinograms generated by different approaches. From left to right: sparse-view sinogram, linearly interpolated sinogram, deep learning synthesized sinogram, and full-view sinogram (reference). The display window for all images is [0, 0.6539].

Table 2. The RMSE of sinograms from linear interpolation approach and deep learning approach when compared with the reference sinogram

| | Linear Interpolation | Deep Learning |
|---|---|---|
| **RMSE** | 0.008087 | **0.004709** |

### 3.2 The evaluation of reconstructed image

All the sinograms were reconstructed by FBP algorithm. The reconstructed images are displayed in Fig. 4. We can visually observe the severe sparse-view artifact in the sparse-view image. For the image reconstructed from linearly interpolated sinogram, the artifact is still obvious even though it is suppressed to some extent. For the image reconstructed from deep learning synthesized sinogram, the image quality is almost comparable with reference image (the image reconstructed from full-view sinogram). The line profiles of the central vertical lines are displayed in Fig. 5. We can also find in the line profile plot that the image from deep learning approach is the closest one to the reference image.

The RMSE and SSIM are reported in Table 3. When using linear interpolation method to process sparse-view sinograms, the RMSE can be improved from 0.009028 to 0.003245, and the SSIM can be improved from 0.8577 to 0.9869, while when using deep learning method to synthesize sinograms, the RMSE can be improved from 0.009028 to 0.001977, and the SSIM can be improved from 0.8577 to 0.9948. Our proposed approach leads to greater image quality improvement.

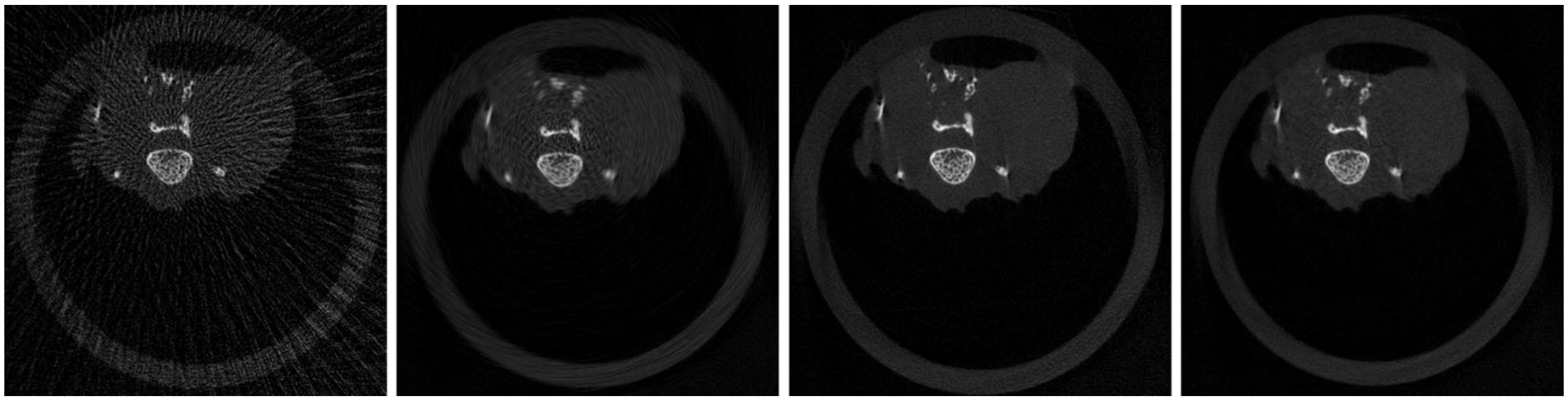

Figure 4. The reconstructed images from different sinograms. From left to right: image from sparse-view sinogram, image from linearly interpolated sinogram, image deep learning synthesized sinogram, and image from full-view sinogram (reference). The display window for all images is $[0,\ 0.1257\ mm^{-1}]$.

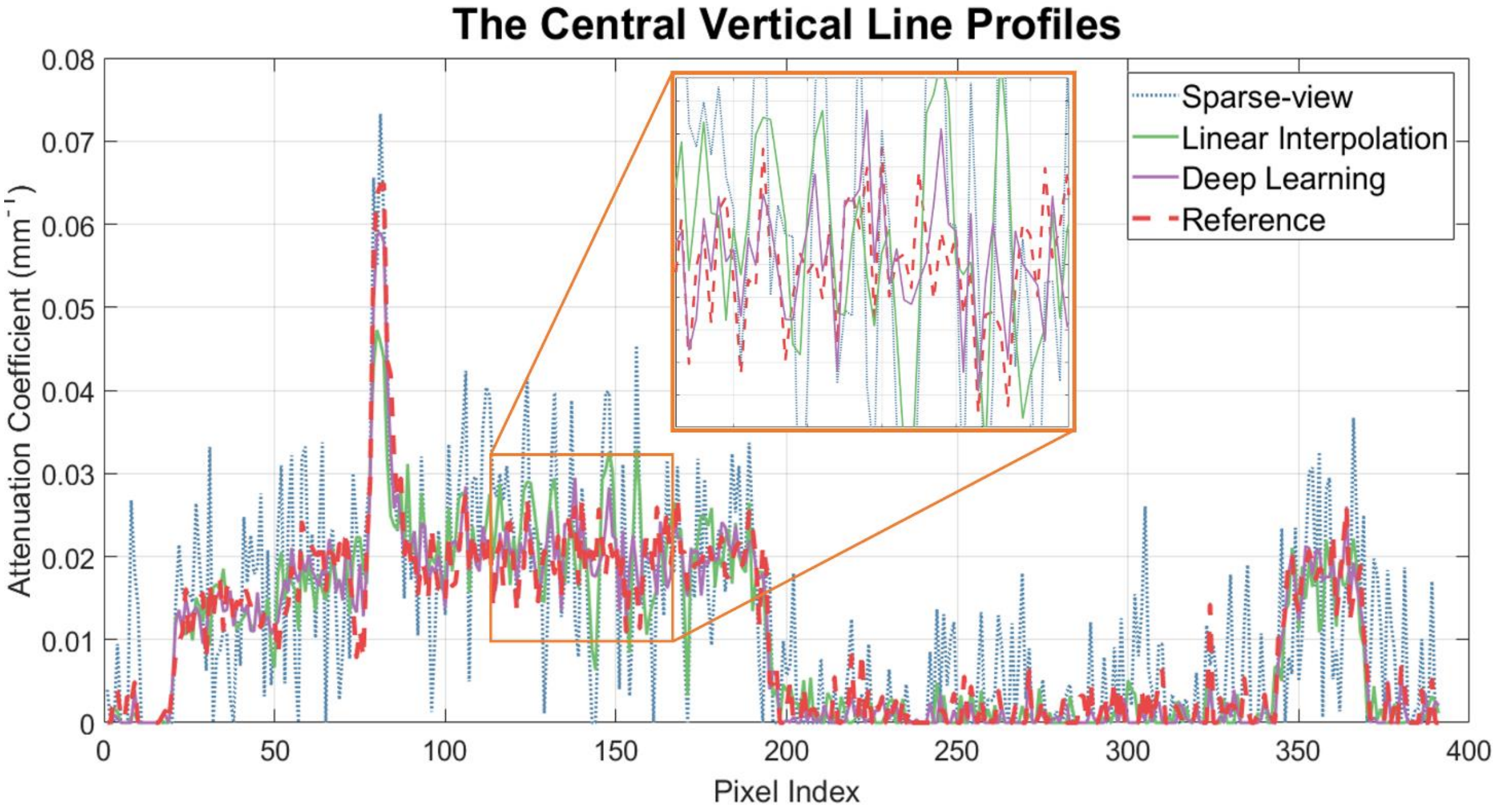

Figure 5. Quantitative line intensity profiles comparison. The line intensity profiles correspond to the central vertical lines in the CT images shown in Fig. 4 reconstructed from the sparse-view sinogram (blue dashed line), linearly interpolated sinogram (green solid line), image from deep learning synthesized sinogram (purple solid line), and image from full-view sinogram (red dashed line).

Table 3. The RMSE and SSIM of images reconstructed from different approaches when compared with reference image

| | Sparse-view | Linear Interpolation | Deep Learning |
|---|---|---|---|
| **RMSE** | 0.009028 | 0.003245 | **0.001977** |
| **SSIM** | 0.8577 | 0.9869 | **0.9948** |

# 4. DISCUSSION AND CONCLUSION

In this study, we developed a sinogram synthesis method for sparse-view micro-CT based on deep learning neural network. The network is based on the combination of U-Net and residual learning. The U-Net can effectively learn the intricate structure of sinogram data especially for biomedical images. The residual learning can facilitate training by learning the difference between the input sinogram data and reference sinogram data instead of directly transforming sinogram data. We applied the method to micro-CT experimental data and the result shows the RMSE was improved from 0.009028 to 0.001977 and SSIM was improved from 0.8577 to 0.9948.

In our study, we first interpolated the sinogram using simple linear interpolation technique and then utilized neural network to further improve the linearly interpolated sinogram. This interpolation process can make sure the data size of the network input is the same as the data size of the network output. Maintaining the network input and output in the same size can make the training easier.

The deep learning based sparse-view sinogram synthesis approaches have been explored by some previous studies [15, 17, 20]. For example, Lee et al. proposed a successive convolutional neural network with 20 layers [17]; Ghani et al. applied conditional GAN to process the sparse-view sinograms [20]. But all those studies use simulation data instead of experimental data. In this study, we applied our approach to micro-CT experimental data and the image quality was largely improved (RMSE improved 78.10%). Besides, in our study the sparse-view CT data is just 10% of the full-view CT data (90% sparsity). Our approach demonstrates good performance when dealing with realistic sinogram data with very large sparsity.

In summary we developed a DL-based method to synthesize sinogram for sparse-view micro-CT. When applying the method to experimental sinogram data with very large sparsity, the image quality was improved significantly. The proposed method could lead to high quality micro-CT imaging with faster imaging speed and lower radiation dose.